# (Accepted Manuscript) A new moment-independent uncertainty importance measure based on cumulative residual entropy for developing uncertainty reduction strategies

Shi-Shun Chen, Xiao-Yang Li *

School of Reliability and Systems Engineering, Beihang University, Beijing 100191, China

Email: css1107@buaa.edu.cn (Shi-Shun Chen), leexy@buaa.edu.cn (Xiao-Yang Li)

Corresponding author*: Xiao-Yang Li

**This manuscript has been accepted by *Mathematics and Computers in Simulation***

Article link: https://www.sciencedirect.com/science/article/pii/S0378475425002253

Code link: https://github.com/dirge1/GSA_CRE

## Highlights

- Quantifying uncertainty magnitude is important in practical uncertainty reduction.
- CRE-based measures are developed for handling highly-skewed distributions.
- Numerical implementations are devised to estimate the proposed measure.
- A case of uncertainty reduction considering uncertainty magnitude is introduced.
- The proposed measure gives a different recommendation compared to the Sobol index.

## Abstract

Uncertainty reduction is crucial for enhancing system reliability and mitigating risks. To identify the most effective target for uncertainty reduction, uncertainty importance measures are commonly used in global sensitivity analysis to prioritize input variable uncertainties. Designers then take steps to reduce the uncertainties of variables with high importance. However, for variables with minimal uncertainty, the cost of controlling their uncertainties can be unacceptable. Therefore, uncertainty magnitude and the corresponding cost for uncertainty reduction should also be considered when developing uncertainty reduction strategies. Although variance-based methods have been developed for this purpose, they rely on statistical moments and face limitations when handling highly-skewed distributions. Additionally, existing moment-independent methods fail to effectively quantify the uncertainty magnitude and cannot fully support the formulation of uncertainty reduction strategies. Motivated by this issue, we propose a new uncertainty importance measure based on cumulative residual entropy. The proposed measure is moment-independent based on cumulative distribution function, enabling it to handle highly-skewed distributions and quantify uncertainty magnitude effectively. Numerical implementations for estimating the proposed measure are devised and validated. The effectiveness of the proposed measure in importance ranking is verified through two numerical examples, comparing it with the Sobol index, delta index, Gaussian kernel-based index and mutual information. Then, a real-world engineering case involving highly-skewed distributions is presented to illustrate the development of uncertainty reduction strategies considering uncertainty importance and magnitude. The results demonstrate that the proposed measure presents a different uncertainty reduction recommendation compared to the variance-based approach due to its moment-independent characteristic. Our code is publicly available at GitHub: https://github.com/dirge1/GSA_CRE.

## 1 Introduction

Uncertainty reduction is crucial for enhancing system reliability and mitigating the risk of hazardous events. In practice, various sources of uncertainty exist, including manufacturing imperfections, external influences, and the inherent complexity of the systems [1-4]. Thus, a reasonable uncertainty reduction strategy is needed to guide the direction of uncertainty control. Global sensitivity analysis (GSA) is a pivotal technique in assessing how variations in model inputs contribute to output uncertainty [5, 6]. Utilizing the results of GSA, the importance of input variable uncertainties can be prioritized, thus determining the most effective targets for uncertainty reduction. Among existing global sensitivity indices, the variance-based Sobol index has gained much popularity [7]. The fundamental idea of the Sobol index involves decomposing the variance of model output into a series of partial variances linked to single or several input variables. The uncertainty importance is then defined as the ratio of the partial variance to the total output variance. The Sobol index has been successfully applied to the development of uncertainty reduction strategies in various fields, such as aircraft design [8], grinder system design [9], and vehicle rollover risk control [10]. Nevertheless, since variance is derived from the second-order moment, it has a limitation in sufficiently describing the uncertainty of highly-skewed or heavy-tailed distributions, which restricts its applications [11, 12].

In contrast to the Sobol index which focuses solely on the second-order moment of a variable, moment-independent methods are more flexible and powerful. The most typical are distance-based GSA methods [13], like the delta index [14] and the PAWN index [15] which regard the change in probability density function (PDF) and cumulative distribution function (CDF) of the output as uncertainty importance, respectively. For instance, Stover et al. [16] employed delta index in the stochastic unit commitment and demonstrated its effectiveness in dimension reduction. Cardoso-Fernández et al. [17] utilized PAWN index to analyze the importance of input variables with respect to the performance of a generator-absorber heat exchange system, contributing to the optimal design of working conditions. Recently, kernel-based global sensitivity methods [18-20] have been proposed, which can be seen as an extension of distance-based GSA methods. Instead of quantifying the distance of original PDF or CDF distributions, kernel-based methods quantify the distance of the embedding of distributions in reproducing kernel Hilbert spaces using maximum mean discrepancy (MMD). In the framework of kernel-based method, the Gaussian kernel is the most commonly employed for moment-independent GSA [18-20]. Additionally, several existing global sensitivity measures have been shown to be special cases of the kernel-based method, with the Sobol index being a prominent example, which can be derived from the quadratic kernel [21].

In addition to the distance-based GSA methods, Shannon entropy from the information theory has also been widely adopted as an alternative moment-independent uncertainty importance measure [22]. Tang et al. [23] first proposed the Shannon entropy-based importance measure and discussed its mathematical properties. Subsequently, Yazdani et al. [24] applied Shannon entropy-based sensitivity index in assessing the influence of ground motion and structural variables on the engineering demand parameters of structures. Besides, based on the framework of Shannon entropy, mutual information and relative entropy were also used for GSA, including applications in watershed modeling [25] and engine block design [26].

Although many moment-independent global sensitivity methods have been developed to address the drawbacks of variance-based methods, they fail to quantify the uncertainty magnitude and, therefore, cannot fully support the formulation of effective uncertainty reduction strategies. In practical applications, uncertainty reduction requires additional resources, and its cost is significantly related to the uncertainty magnitude [27]. Despite controlling the input variable with the highest uncertainty importance theoretically yields the greatest benefit, the cost of reducing its uncertainty can be unacceptable if its uncertainty magnitude is already sufficiently small. In such cases, controlling the uncertainty of other input variables with high importance could provide greater benefits given cost constraints. Besides, if the uncertainty magnitude of the output variable can be quantified, analysts can access the current level of uncertainty, which is beneficial to establish a threshold demonstrating that uncertainty is controlled within acceptable limits [28]. Consequently, to develop effective uncertainty reduction strategies, it is essential to quantify both the magnitude and importance of input variable uncertainties. However, existing moment-independent GSA indices face challenges in quantifying uncertainty magnitude. Distance-based GSA indices, such as the delta index and the PAWN index, focus solely on the distribution variations and are therefore unable to quantify uncertainty magnitude. For the kernel-based method, Da Veiga [20] developed an uncertainty decomposition framework using MMD, enabling the quantification of the total uncertainty of a random variable in addition to the uncertainty importance. However, for the existing Gaussian kernel-based method [18-20], the calculated total uncertainty approaches unity for any random variable. Therefore, this method also fails to quantify uncertainty magnitude properly. On the other hand, Shannon entropy is suitable for quantifying the uncertainty of discrete random variables based on their PDF. Nonetheless, for continuous random variables, which are more common in real-world applications, differential entropy (the continuous form of Shannon entropy) fails to represent the uncertainty magnitude effectively. For example, in the case of a standard normal distribution with a small variance, the differential entropy can be negative [29], which lacks practical meaning in engineering applications.

Table 1 summarizes the characteristics of the aforementioned global sensitivity methods. It highlights the lack of moment-independent global sensitivity indices capable of quantifying both uncertainty importance and magnitude of continuous variables simultaneously. As a result, it is still challenging in providing effective guidance for developing uncertainty reduction strategies when faced with highly-skewed distributions.

Table 1 Characteristics of the existing global sensitivity methods.

| Global sensitivity methods | Uncertainty importance quantification | Uncertainty magnitude quantification | Moment-independent |
|---|---|---|---|
| Sobol index | √ | √ | |
| Delta index | √ | | √ |
| PAWN index | √ | | √ |
| Kernel-based measure | √ | | √ |
| Shannon entropy-based measure | √ | √ (only applicable to discrete variables) | √ |

To address the above challenge, we introduce the cumulative residual entropy (CRE) [30], which is defined similarly to Shannon entropy. However, unlike Shannon entropy, CRE measures uncertainty based on the CDF rather than the PDF This distinction allows CRE to maintain a unified characteristic for both discrete and continuous variables [31]. Therefore, CRE is able to quantify the uncertainty magnitude of continuous variables legitimately unlike Shannon entropy. Besides, since CRE relies on the CDF rather than statistical moments, the proposed measure is moment-independent and can handle highly-skewed distributions properly. To the best of our knowledge, this is the first time CRE is introduced in developing global sensitivity indices.

The main contributions of this paper can be summarized as follows:

- A new moment-independent uncertainty importance measure based on CRE is proposed, enabling the simultaneous quantification of uncertainty importance and magnitude to support the development of effective uncertainty reduction strategies.
- Numerical implementations are devised and validated to estimate the proposed CRE-based measure.
- A real-world engineering case with a highly-skewed distribution is presented to demonstrate the procedure of developing uncertainty reduction strategies while considering both uncertainty importance and magnitude, which indicates the superiority of the CRE-based measure over existing GSA methods.

The organization of the paper is as follows. The preliminaries of CRE are introduced in Section 2. Next, the CRE-based uncertainty importance measure and corresponding numerical implementations are developed in Section 3. After that, the efficacy of the proposed measure in quantifying uncertainty importance is verified by two numerical examples in Section 4. Subsequently, the proposed measure is applied to an engineering case in Section 5 to show its superiority on developing uncertainty reduction strategies. Finally, Section 6 concludes the work.

## 2 Preliminary

We first recall the definition of differential entropy. According to the information theory, given a continuous random variable $X$, the differential entropy of $X$ can be expressed as [29]:

$$h(X) = -\int_{-\infty}^{+\infty} f(x) \ln f(x) \mathrm{d}x \tag{1}$$

where $f(x)$ is the PDF of $X$.

To show the flaw of differential entropy mentioned in the introduction, we consider an example of a uniform distribution with PDF given by:

$$f(x)=\begin{cases}\frac{x}{b-a}, & a\le x\le b\\ 0, & else\end{cases} \tag{2}$$

Then, the differential entropy of the uniform distribution can be derived as:

$$h(X)=-\int_a^b \frac{1}{b-a}\ln\left(\frac{1}{b-a}\right)\mathrm{d}x=-\ln\left(\frac{1}{b-a}\right)=\ln(b-a). \tag{3}$$

For a uniform distribution $U(0, 0.5)$, from Eq. (3), its differential entropy is -ln2. A negative uncertainty magnitude would imply a more certain state than absolute certainty, which is illogical. At most, uncertainty can be zero, indicating no randomness (e.g., a perfectly deterministic system). Therefore, using negative values to represent uncertainty magnitude is impractical, which reveals the flaw of differential entropy and motivates us to introduce CRE.

In the framework of the information theory, Rao et al. [30] defined the CRE in order to unify the properties of the uncertainty quantification in discrete and continuous random variables. Let $X$ be a nonnegative random variable, the CRE of $X$ is defined by:

$$\mathcal{E}(X):=-\int_0^{+\infty}\bar{F}(x)\ln\left[\bar{F}(x)\right]\mathrm{d}x \tag{4}$$

where $\bar{F}(x)=P(X>x)=1-F(x)$ is the cumulative residual function or survival function of $X$, and $F(x)$ is the CDF of $X$. It is worth noting that Rao only defined CRE for nonnegative random variables. Later, Drissi et al. [32] extended the definition of $\mathcal{E}(X)$ to the case with support in $\mathbb{R}$. Then, Eq. (4) can be rewritten as:

$$\mathcal{E}(X):=-\int_{-\infty}^{+\infty}\bar{F}(x)\ln\left[\bar{F}(x)\right]\mathrm{d}x \tag{5}$$

To illustrate the analytical calculations of CRE clearly, we present three examples in the following.

***Example 1***: The exponential distribution with mean $1/\lambda$ has the CDF:

$$F(x)=\begin{cases}1-e^{-\lambda x}, x\ge 0\\ 0, x<0\end{cases} \tag{6}$$

Correspondingly, the CRE of the exponential distribution is:

$$\begin{aligned}\mathcal{E}(X)&=-\int_0^{+\infty}e^{-\lambda x}\ln\left(e^{-\lambda x}\right)\mathrm{d}x=\int_0^{+\infty}\lambda x e^{-\lambda x}\mathrm{d}x\\ &=\int_0^{+\infty}e^{-\lambda x}\mathrm{d}x=\frac{1}{\lambda}\end{aligned} \tag{7}$$

***Example 2***: The uniform distribution has the CDF:

$$F(x)=\begin{cases}\frac{x-a}{b-a}, & a\le x\le b\\ 0, & else\end{cases} \tag{8}$$

Let $u=\frac{b-x}{b-a}$, then the CRE of the uniform distribution can be deduced as:

$$\mathcal{E}(X)=-\int_a^b\left(\frac{b-x}{b-a}\right)\ln\left(\frac{b-x}{b-a}\right)\mathrm{d}x=(b-a)\int_1^0 u\ln u\mathrm{d}u=\frac{b-a}{4} \tag{9}$$

***Example 3***: The Gaussian distribution has the CDF:

$$F\left(x\right)=1-\text{erfc}\left(\frac{x-\mu}{\sigma}\right) \tag{10}$$

where erfc is the error function:

$$\text{erfc}(x)=\frac{1}{\sqrt{2\pi}}\int_{x}^{+\infty}\exp\left(-\frac{t^2}{2}\right)dt \tag{11}$$

Let $u=\frac{x-\mu}{\sigma}$, then the CRE of the Gaussian distribution can be calculated as:

$$\mathcal{E}\left(X\right)=-\int_{-\infty}^{+\infty}\text{erfc}\left(\frac{x-\mu}{\sigma}\right)\ln\left(\text{erfc}\left(\frac{x-\mu}{\sigma}\right)\right)\mathrm{d}x=-\sigma\int_{-\infty}^{+\infty}\text{erfc}\left(u\right)\ln\left(\text{erfc}\left(u\right)\right)\mathrm{d}u\approx 0.9032\sigma \tag{12}$$

For the uniform distribution $U(0, 0.5)$ mentioned above, from Eq. (9), its CRE is 0.125, which demonstrates why CRE is a more suitable measure for quantifying uncertainty magnitude compared to differential entropy.

## 3 Cumulative residual entropy-based uncertainty importance measure

To facilitate the description of the proposed measure, we introduce the following generic model. It posits that the output response of a system is determined by numerous input random variables, which can be expressed as follows:

$$Y=g\left(\boldsymbol{X}\right), \tag{13}$$

where $\boldsymbol{X}=(X_1, X_2, \cdots, X_n)$ is an $n$-dimensional vector of the input variables with uncertainties; $Y$ represents the output response with uncertainty propagated by $\boldsymbol{X}$ via the function $g(\cdot)$. In the following sections, we propose the uncertainty importance measure in the framework of CRE based on Eq. (13).

According to existing research [7, 14, 23], the contribution of an input variable's uncertainty to output uncertainty can be divided into two components: isolated contribution and interactive contribution with other variables. This principle is widely recognized across various methods and aligns with practical applications. Therefore, two types of CRE-based uncertainty importance measures are introduced in this section. The CRE-based uncertainty importance measure for a single variable is introduced in Section 3.1, quantifying the isolated contribution, and the CRE-based uncertainty importance measure for interactions between variables is introduced in Section 3.2, capturing the interactive contribution.

**Remark 1**: When input variables are correlated, the contributions of input uncertainty to output uncertainty become more complex [33-35]. Correlation reduces the isolated contribution while also influencing interaction contributions. In this paper, we focus solely on independent input variables. Quantifying uncertainty importance based on CRE for correlated variables remains an open challenge and will be explored in future research.

### 3.1 CRE-based uncertainty importance measure for single variable

Firstly, based on the framework of CRE, we can define the uncertainty importance measure of a single variable as:

$$\kappa_i := \frac{\mathcal{E}\left(Y\right)-E_{X_i}\left[\mathcal{E}\left(Y|X_i=x_i\right)\right]}{\mathcal{E}\left(Y\right)}=1-\frac{E_{X_i}\left[\mathcal{E}\left(Y|X_i=x_i\right)\right]}{\mathcal{E}\left(Y\right)} \tag{14}$$

where $\mathcal{E}\left(Y\right)$ is the CRE of the output variable $Y$, which is formulated by Eq. (5); $E_{X_i}\left[\cdot\right]$ represents the expectation taken with respect to $X_i$; and $\mathcal{E}(Y|X_i = x_i)$ is the conditional CRE of $Y$ when $X_i = x_i$, which is defined by:

$$\mathcal{E}(Y|X_i = x_i) = -\int_{-\infty}^{+\infty} \bar{F}(Y|X_i = x_i) \ln\left[\bar{F}(Y|X_i = x_i)\right] \mathrm{d}y \tag{15}$$

where $\bar{F}(Y|X_i = x_i)$ denotes the survival function of $Y$ when $X_i = x_i$. From Eq. (15), $\mathcal{E}(Y|X_i = x_i)$ quantifies the remaining uncertainty of $Y$ when the uncertainty of $X_i$ is removed by fixing its value at $x_i$, and $E_{X_i}\left[\mathcal{E}(Y|X_i = x_i)\right]$ quantifies the expectation of remaining uncertainty of $Y$ when the uncertainty of $X_i$ is removed. Consequently, the importance measure formulated by Eq. (14) can quantify the isolated contribution of single input variable uncertainty to the output uncertainty.

### 3.2 CRE-based uncertainty importance measure for interactions between variables

In addition to the uncertainty importance of single variable, uncertainty importance between variables is also significant to understand their interaction contributions. To define this measure, we first draw on the definition of mutual information from differential entropy. The cumulative residual mutual information (CRMI) between $X$ and $Y$ can be expressed as:

$$I_{\mathcal{E}}(X;Y) = \mathcal{E}(Y) - E_X\left[\mathcal{E}(Y|X = x)\right] \tag{16}$$

Besides, for any random variable $X$, $Y$ and $Z$, the multivariate cumulative residual mutual information (MCRMI) between $X$, $Y$ and $Z$ can be described as:

$$I_{\mathcal{E}}(X,Y;Z) = \mathcal{E}(Z) - E_{X,Y}\left[\mathcal{E}(Z|X = x, Y = y)\right] \tag{17}$$

where $\mathcal{E}(Z|X = x, Y = y)$ represents the conditional CRE of $Z$ when $X = x$ and $Y = y$. $I_{\mathcal{E}}(X;Y)$ gives the expectation of uncertainty reduction of $Y$ given $X$, while $I_{\mathcal{E}}(X,Y;Z)$ gives the expectation of uncertainty reduction of $Z$ given $X$ and $Y$.

Then, based on Eqs. (16) and (17), for two independent variables $X_i$ and $X_j$, their interaction contributions to the uncertainty of $Y$ can be defined by:

$$\kappa_{ij} = \frac{I_{\mathcal{E}}(X_i, X_j;Y) - I_{\mathcal{E}}(X_i;Y) - I_{\mathcal{E}}(X_j;Y)}{\mathcal{E}(Y)} \tag{18}$$

The importance measure formulated by Eq. (18) can quantify the interaction contributions of two independent input variables to the output uncertainty in addition to their isolated contributions.

**Remark 2**: Based on Eq. (18), the interaction contribution among more independent variables can be given in a similar way. To be specific, the interaction contributions of all the input variables can be derived as:

$$\kappa_{12\ldots n} = \frac{I_{\mathcal{E}}(X_1, X_2, \ldots, X_n;Y)}{\mathcal{E}(Y)} - \sum_{i=1}^{n} \kappa_i - \sum_{1 \le i < j \le n} \kappa_{ij} - \ldots - \sum_{1 \le i < j \le \ldots \le k \le n} \kappa_{ij\ldots k} \tag{19}$$

where

$$I_{\mathcal{E}}(X_i, X_j, \ldots, X_n;Y) = \mathcal{E}(Y) - E_{X_i, X_j, \ldots, X_n}\left[\mathcal{E}(Y|X_i = x_i, X_j = x_j, \ldots, X_n = x_n)\right] \tag{20}$$

### 3.3 Properties of the CRE-based uncertainty importance measure

In order to investigate the properties of the proposed uncertainty importance measure formulated by Eqs. (14) and (18), we first introduce several theorems of CRE.

***Theorem 1*** *(see [32])*: For any random variable *X*,

$$\mathcal{E}(X) \geq 0 \tag{21}$$

and equality holds if and only if *X* is a constant. It should be noted that this theorem is not satisfied for the Shannon entropy of continuous variables as shown in Section 2.

***Definition*** *(see [30])*: A sigma field $\mathcal{F}$ is a class of subsets containing the empty set and closed under compliments and countable unions.

***Theorem 2*** *(see [32])*: For any random variables *X* and a given sigma field $\mathcal{F}$, which represents a collection of measurable events that may contain partial information about *X*, we have:

$$\mathcal{E}(X|\mathcal{F}) \leq \mathcal{E}(X) \tag{22}$$

and equality holds if and only if *X* is independent of $\mathcal{F}$. *X* is said to be independent of the sigma field $\mathcal{F}$ indicates that it is independent of any random variable that can be measured with respect to $\mathcal{F}$. If $\mathcal{F}$ is the sigma field generated by a random variable *Y*, then we have:

$$E_Y\left[\mathcal{E}(X|Y=y)\right] \leq \mathcal{E}(X) \tag{23}$$

and equality holds if and only if *X* is independent of *Y*.

***Theorem 3*** *(see [32])*: For any random variable *X* and a given sigma field $\mathcal{F}$, we have:

$$\mathcal{E}(X|\mathcal{F}) = 0 \ \text{ iif } \ X \ \text{ is } \mathcal{F} \text{ measurable.} \tag{24}$$

If $\mathcal{F}$ is the sigma field generated by a random variable *Y*, then we have:

$$E_Y\left[\mathcal{E}(X|Y=y)\right] = 0 \ \text{ iif } \ X \ \text{ is a function of } Y. \tag{25}$$

Subsequently, inspired by the properties of delta index discussed in [14], this paper presents the following four properties for the CRE-based uncertainty importance measure.

***Property 1***. $0 \leq \kappa_i \leq 1$.

*Proof*: The proof is in the Appendix A.1.

***Property 2***. If the output *Y* is independent of $X_i$, then $\kappa_i = 0$; if *Y* is a function of $X_i$, then $\kappa_i = 1$.

*Proof*: The former statement can be easily proved by the equality condition of *Theorem 2*. The latter one can be verified by *Theorem 3*. □

***Property 3***. $0 \leq \kappa_{ij} \leq 1$.

*Proof*: The proof is in the Appendix A.2.

***Property 4***. $\sum_{i=1}^{n}\kappa_i + \sum_{1\leq i<j\leq n}\kappa_{ij} + \ldots + \kappa_{12\ldots n} = 1$.

*Proof*: The proof is in the Appendix A.3.

In summary, the $\kappa$ of a single variable or interactions between variables ranges from 0 to 1. The importance of variables independent of the output is zero. The total importance of all variables sums to unity. These four properties ensure that the output uncertainty can be reasonably decomposed to isolated and interacted contributions of the input variables. Besides, Theorem 1 guarantees that the uncertainty magnitude can be legitimately quantified using CRE, i.e., the uncertainty magnitude is greater than or equal to zero for any continuous variables. Thus, the CRE-based measure is able to quantify the uncertainty importance and uncertainty magnitude simultaneously for developing effective uncertainty reduction strategies.

**Remark 3**: It should be noted that the partial information decomposition theory mentioned in the proof of Property 3 has a similar framework as variance decomposition [34]. This suggests that the interaction contribution quantified through CRE corresponds to the same type of uncertainty as in variance analysis (i.e., ANOVA). Nevertheless, because of their distinct theoretical foundations, they present different analytical results.

**Remark 4**: Importantly, the uncertainty magnitude of an input does not necessarily reflect its uncertainty importance. For example, an input $X_1$ may have a small uncertainty magnitude but high uncertainty importance, while another input $X_2$ may have a moderate uncertainty magnitude yet low importance. The purpose of quantifying uncertainty magnitude is to reflect the cost implications in practical uncertainty reduction scenarios, which is non-relevant to the assessment of uncertainty importance.

### 3.4 Estimation of cumulative residual entropy-based importance measure

In this section, numerical estimations are presented for the CRE-based uncertainty importance measure. In order to compute the importance measure of a single variable formulated by Eq. (14), the estimations of CRE and conditional CRE given single variable are required, which are developed in Sections 3.4.1 and 3.4.2, respectively. Then, in order to calculate the measure of interaction contributions formulated by (18), the estimation of conditional CRE given two variables is needed, which is presented in Section 3.4.3.

#### 3.4.1 Estimation of cumulative residual entropy

Rao et al. [30] developed an empirical estimator of CRE based on the empirical CDF. Given a set of $n$ samples from random variable $X$, denoted as $\left(x_1, x_2, \ldots, x_n\right)$, the empirical estimator of CRE is defined as:

$$\hat{\mathcal{E}}\left(X\right) = -\sum_{i=1}^{n-1} U_{i+1}\left(1-\frac{i}{n}\right)\log\left(1-\frac{i}{n}\right), \tag{26}$$

where $U_i$ is the sample spacing denoted by

$$U_1 = x_{(1)}, U_i = x_{(i)} - x_{(i-1)}, i = 2,3,\ldots,n, \tag{27}$$

where $x_{(1)} < x_{(2)} < \ldots < x_{(n)}$ denotes the order statistics of the samples.

In Eq. (26), the integral formulation of CRE in Eq. (5) is discretized by replacing the continuous domain with the ordered sample values. The sample spacing $U_i$ approximates the small interval of integration, and ensures that the contribution of each interval is properly weighted. The term $1-i/n$ represents the empirical survival function at the $i$-th order statistic. The convergence validation and computational efficiency of the estimation method by Eq. (27) are given in Appendix B.1.

#### 3.4.2 Estimation of conditional CRE given single variable

Given a set of $n$ samples from a two-dimensional random variable $\left(X,Y\right)$, denoted as $\left[\left(x_1, y_1\right), \left(x_2, y_2\right), \ldots, \left(x_n, y_n\right)\right]$, the samples of $X$ are first sorted in ascending order to obtain the order statistics $x_{(1),r_1} < x_{(2),r_2} < \ldots < x_{(n),r_n}$, where $r_i$ represents the original index of each sorted sample of $X$.

The ordered values of $X$ are then divided into $n/m$ groups, with each group containing $m$ samples. Subsequently, the expectation of conditional CRE given single variable can be estimated as:

$$E_X\left[\mathcal{E}\left(Y|X=x\right)\right]\approx\frac{m}{n}\sum_{i=0}^{n/m-1}\hat{\mathcal{E}}\left(Y_i\right) \tag{28}$$

where $Y_i=\left(y_{r_{im+1}},y_{r_{im+2}},\dots,y_{r_{im+m}}\right)$ denotes the samples of $Y$ belonging to the $(i+1)$-*th* group.

The idea behind Eq. (28) is to discretize the data while preserving the conditional structure of $Y$ given $X$. Discretization is necessary because the empirical conditional CDF cannot be directly obtained, making it impossible to compute the integral required for conditional CRE. To achieve this, the order statistics of $X$ are divided into equal-sized groups, transforming $X$ from a continuous variable into a discrete one with $n/m$ possible values, each occurring with a probability of $m/n$. Next, the conditional CRE of $Y$ can be estimated by first computing the empirical CRE of $Y$ in each discrete group, and then taking a probability-weighted sum of these values to obtain the final estimate. The convergence validation and computational efficiency of the estimation method by Eq. (28) are given in Appendix B.2.

### 3.4.3 Estimation of conditional CRE given two variables

Given a set of $n$ samples from a three-dimensional random variable $\left(X,Y,Z\right)$, denoted as $\left[\left(x_1,y_1,z_1\right),\left(x_2,y_2,z_2\right),\dots,\left(x_n,y_n,z_n\right)\right]$, the samples of $X$ and $Y$ are first sorted to obtain their order statistics $x_{(1),r_1}<x_{(2),r_2}<\dots<x_{(n),r_n}$ and $y_{(1),s_1}<y_{(2),s_2}<\dots<y_{(n),s_n}$, where $s_i$ represents the original index of each sorted sample of $Y$. Next, the ordered samples of $X$ and $Y$ are partitioned into $I$ and $J$ groups, with each group containing $n/I$ and $n/J$ samples, respectively. Then, the expectation of conditional CRE given two variables can be estimated as:

$$E_{X,Y}\left[\mathcal{E}\left(Z|X,Y\right)\right]\approx\sum_{i=1}^{I}\sum_{j=1}^{J}\frac{N_{ij}}{n}\hat{\mathcal{E}}\left(Z_{ij}\right), \tag{29}$$

where $N_{ij}$ represents the number of samples of $Z$ that falls into the $i$-th group of $X$ and the $j$-th group of $Y$; and $Z_{ij}=\left\{z_{r_{\frac{in}{I}+1}},z_{r_{\frac{in}{I}+2}},\dots,z_{r_{\frac{in}{I}+\frac{n}{I}}}\right\}\cap\left\{z_{s_{\frac{jn}{J}+1}},z_{s_{\frac{jn}{J}+2}},\dots,z_{s_{\frac{jn}{J}+\frac{n}{J}}}\right\}$ denotes the samples of $Z$ belonging to the $i$-th group of $X$ and the $j$-th group of $Y$.

The idea behind Eq. (29) is similar to that of Eq. (28), involving data discretization while preserving the conditional structure of $Z$ given $X$ and $Y$. The order statistics of $X$ and $Y$ are divided into groups of equal sample size, transforming them from continuous variables into discrete ones with $I$ and $J$ possible values, respectively. This transformation creates a two-dimensional probability space of $X$ and $Y$ with $I\times J$ discrete variable combinations, and the probability of the discrete variable corresponding to the $i$-th group of $X$ and the $j$-th group of $Y$ is $N_{ij}/n$. Next, the conditional CRE of $Z$ is estimated by first computing the empirical CRE of $Z$ in each discrete combination of $X$ and $Y$, and then performing a probability-weighted summation to obtain the final estimate. The convergence validation and computational efficiency of the estimation method by Eq. (29) are given in Appendix B.3.

## 4 Numerical examples

In this section, we show the effectiveness of the CRE-based measure on quantifying uncertainty importance through two numerical examples. The CRE-based importance measure is denoted as $\kappa$. Five other GSA indices are chosen for comparison: main effect ($S$) and total effect ($S^T$) based on the variance-based Sobol index [7], delta index ($\delta$) [14], Gaussian kernel-based method using MMD ($\phi$) [20], and mutual information based on Shannon entropy ($\eta$) [22]. For the hyper-parameters in the numerical estimation of the CRE-based measure, we set $m = 500$ and $I = J = 20$. The basis for determining the hyper-parameters can be found in Appendix C. The calculation is executed on 11th Gen Intel(R) Core(TM) i7-11800H @ 2.30GHz laptop with 16 GB RAM.

### 4.1 Example 1: Ishigami test function

Consider the following Ishigami test function [36]:

$$Y = \sin X_1 + a \sin^2 X_2 + b X_3^4 \sin X_1 \tag{30}$$

where $X_p \sim U(-\pi,\pi)$, $p = 1, 2, 3$. This nonlinear non-monotonic function is often used in the literature as a benchmark for GSA methods [37]. Here we set $a = 5$ and $b = 1$.

Sensitivity analysis is performed on the output $Y$. Fig. 1 shows the convergence of CRE-based measure for the Ishigami test function. All the CRE-based GSA indices of single variable are converged over 20000 samples, and the result is also satisfactory under 10000 samples. As for the interaction contributions, they get convergence over 40000 samples.

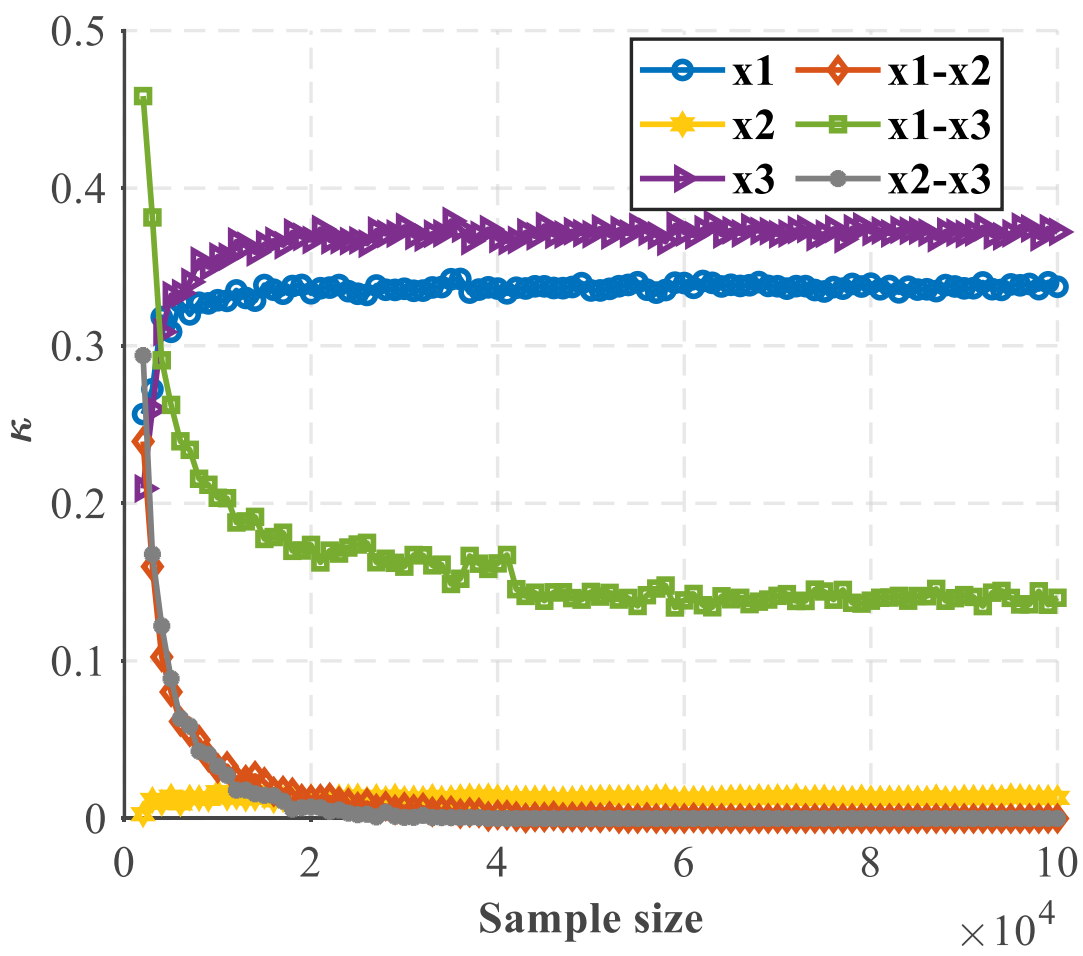

Fig. 1 Convergence of the CRE-based GSA measure of single input variable and interaction contributions for the Ishigami test function.

Table 2 displays the sensitivity results obtained by different GSA methods. As for the importance ranking, all the moment-independent methods support that $X_3$ is the most influential variable, followed by $X_1$, while $X_2$ plays a minor role. However, for the variance-based method, it supports that $X_1$ is the most influential variable. This proves that the CRE-based measure can give a credible importance ranking as well as other moment-independent methods.

Table 2 Global sensitivity results obtained by different GSA methods for the Ishigami test function.

| Variable | $S$ | $S^T$ | $\delta$ | $\phi$ | $\eta$ | $\kappa$ |
|---|---|---|---|---|---|---|
| $X_1$ | 0.3813 (1) | 0.9950 (1) | 0.3394 (2) | 0.0719 (2) | 0.6082 (2) | 0.3381 (2) |
| $X_2$ | 0.0057 (2) | 0.0057 (3) | 0.1325 (3) | 0.0025 (3) | 0.1704 (3) | 0.0129 (3) |
| $X_3$ | 0.0008 (3) | 0.6131 (2) | 0.5096 (1) | 0.1109 (1) | 0.8823 (1) | 0.3734 (1) |

*Note*: The number shown in parentheses denotes the importance ranking of a variable for a given GSA index.

Then, the uncertainty decomposition result for the Ishigami test function obtained by the CRE-based measure is exhibited in Fig. 2. It can be seen that the interaction contributions of $\{X_1, X_2\}$ and $\{X_2, X_3\}$ are zero, which is consistent with the real contributions formulated by Eq. (30).

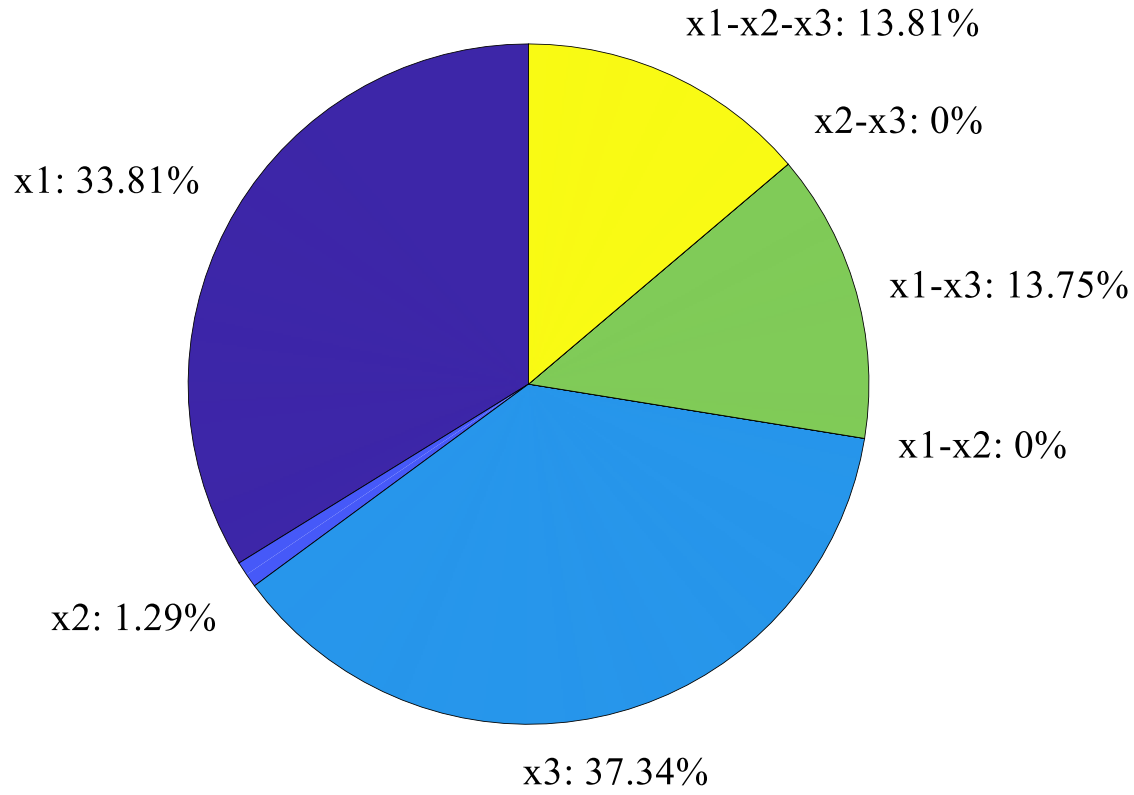

Fig. 2 Uncertainty decomposition results of the CRE-based measure for Ishigami test function.

## 4.2 Example 2: A probabilistic risk assessment model

We then introduce a probabilistic risk assessment model to verify the proposed measure. The Boolean expression of the top event is [38]:

$$\begin{aligned} Y = & X_1X_3X_5 + X_1X_3X_6 + X_1X_4X_5 + X_1X_4X_6 + X_2X_3X_4 \\ & + X_2X_3X_5 + X_2X_4X_5 + X_2X_5X_6 + X_2X_4X_7 + X_2X_6X_7 \end{aligned} \tag{31}$$

where $X_1$ and $X_2$ are initiating events indicating the number of occurrences per year, and $X_3$–$X_7$ are basic events which represent different component failure rate. All the input variables are independent and follow lognormal distributions, as indicated by the results in [38]. Their distribution parameters are listed in Table 3.

Table 3 Distribution parameters of the risk assessment model.

| Variable | Distribution | Mean | Error factor |
|---|---|---|---|
| $X_1$ | Lognormal | 2 | 2 |
| $X_2$ | Lognormal | 3 | 2 |
| $X_3$ | Lognormal | 0.001 | 2 |
| $X_4$ | Lognormal | 0.002 | 2 |
| $X_5$ | Lognormal | 0.004 | 2 |
| $X_6$ | Lognormal | 0.005 | 2 |
| $X_7$ | Lognormal | 0.003 | 2 |

Sensitivity analysis is conducted with respect to the top event. Fig. 3 illustrates the convergence of the CRE-based measure. From Fig. 3 (a), all the CRE-based measures of single variable are converged over 20000 samples, and the result is also satisfactory under 10000 samples. Then, as can be seen in Fig. 3 (b), the CRE-based measures of the interaction contributions containing $X_1$ get convergence over 40000 samples.

Table 4 shows the sensitivity results obtained by different GSA methods. As for the importance ranking, all the GSA methods support that $X_1$ is the most influential variable, followed by $X_6$ and $X_5$, while $X_4$, $X_7$, $X_1$ and $X_3$ play minor roles. This proves that the CRE-based measure is able to give a reliable importance ranking as well as other methods.

Then, the uncertainty decomposition result for the probabilistic risk assessment model obtained by the CRE-based measure is exhibited in Fig. 4. There exists interaction contributions between almost every two variables. Since the interaction of the model is complex, the higher-order contributions are high, with the figure around 30.85%.

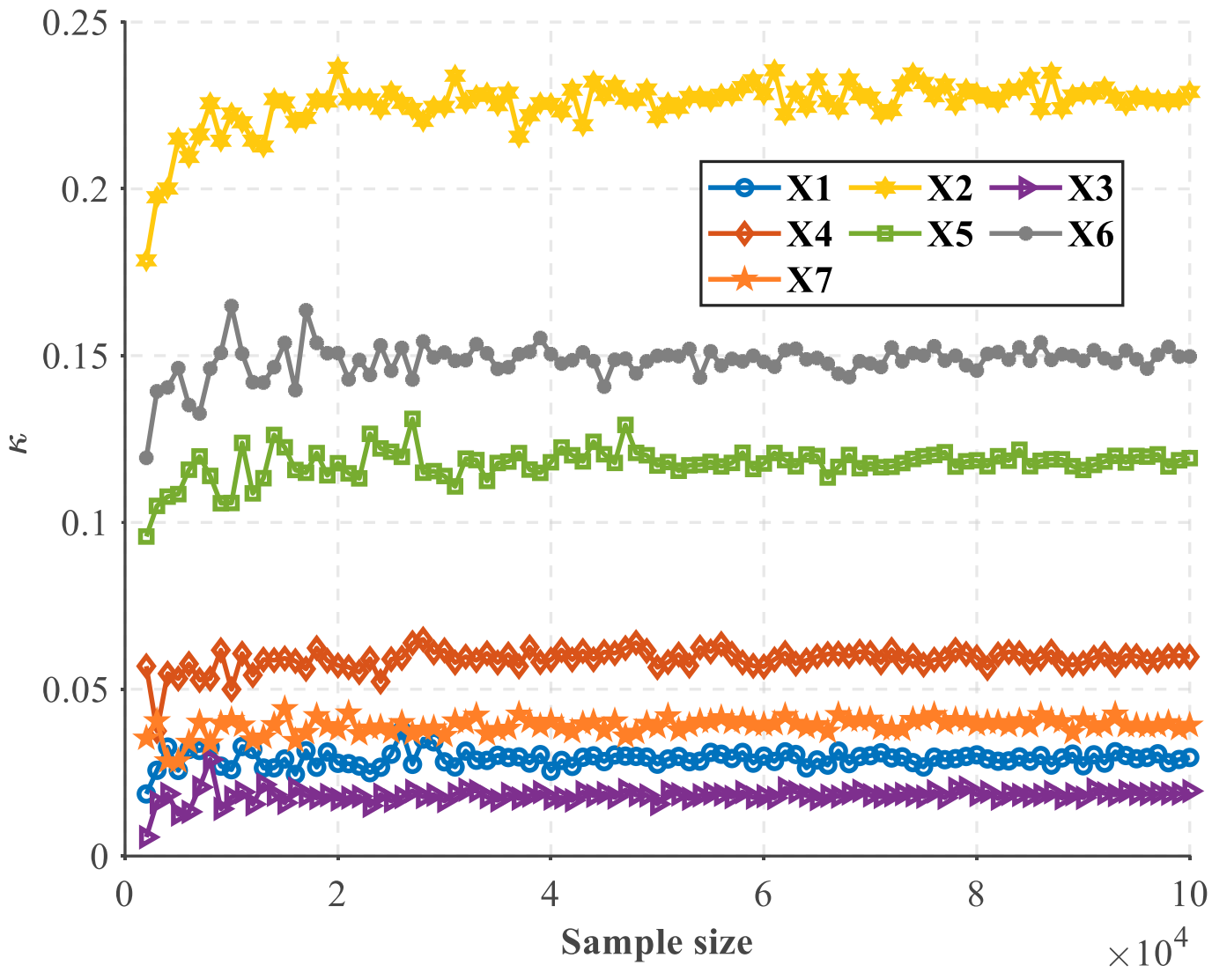

(a)

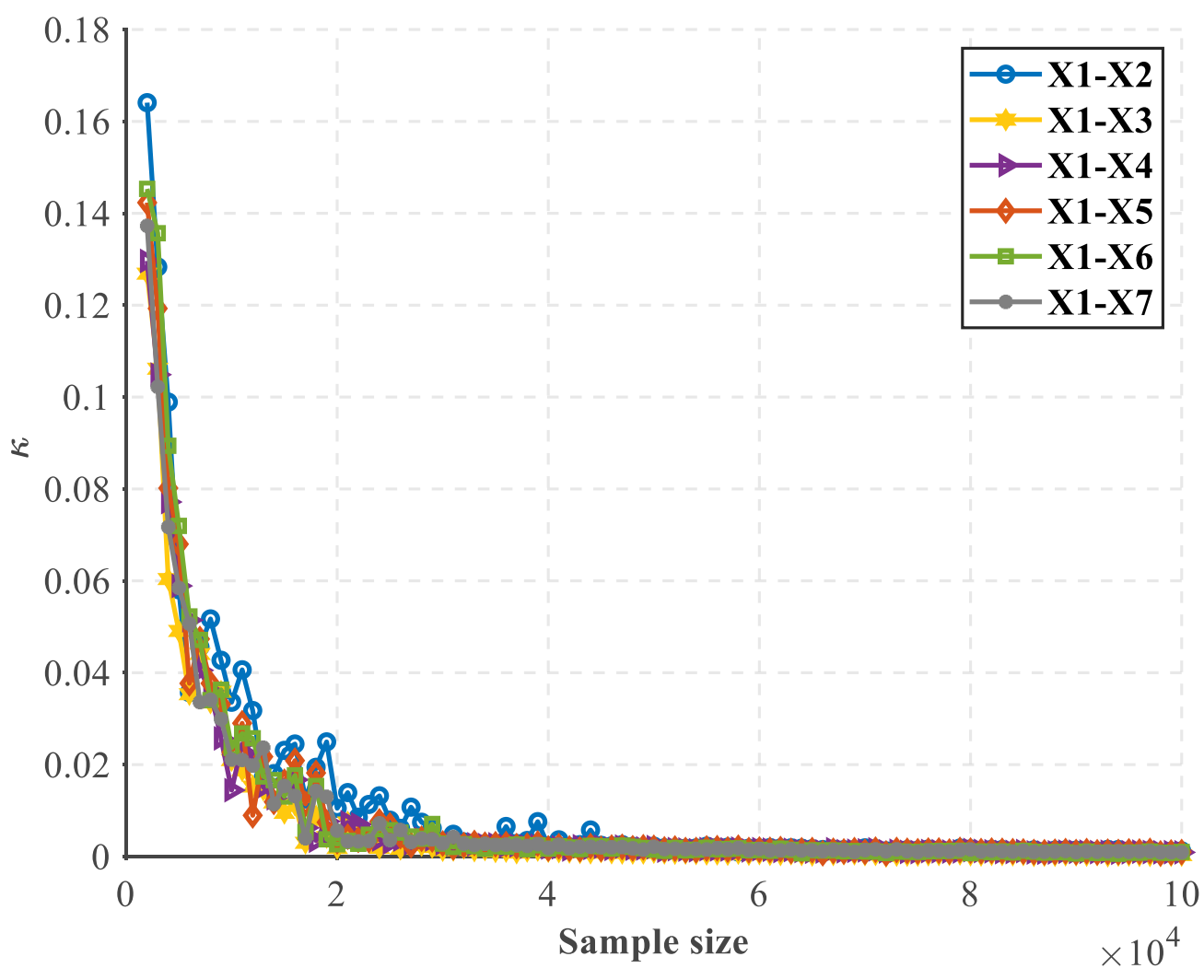

(b)

Fig. 3 Convergence of CRE-based measure of single input variable and interaction contributions

containing $X_1$ for the probabilistic risk assessment model.

Table 4 Global sensitivity results obtained by different GSA methods for the probabilistic risk assessment model.

| Variable | $S$ | $S^T$ | $\delta$ | $\phi$ | $\eta$ | $\kappa$ |
|---|---|---|---|---|---|---|
| $X_1$ | 0.0353 (6) | 0.0428 (6) | 0.0707 (6) | 0.0082 (6) | 0.0221 (6) | 0.0294 (6) |
| $X_2$ | 0.3286 (1) | 0.3953 (1) | 0.2024 (1) | 0.0689 (1) | 0.2139 (1) | 0.2240 (1) |
| $X_3$ | 0.0157 (7) | 0.0186 (7) | 0.0574 (7) | 0.0036 (7) | 0.0150 (7) | 0.0195 (7) |
| $X_4$ | 0.0852 (4) | 0.0998 (4) | 0.1011 (4) | 0.0184 (4) | 0.0600 (4) | 0.0589 (4) |
| $X_5$ | 0.1741 (3) | 0.2124 (3) | 0.1444 (3) | 0.0348 (3) | 0.0998 (3) | 0.1213 (3) |
| $X_6$ | 0.2197 (2) | 0.2654 (2) | 0.1623 (2) | 0.0447 (2) | 0.1408 (2) | 0.1480 (2) |
| $X_7$ | 0.0476 (5) | 0.0638 (5) | 0.0761 (5) | 0.0095 (5) | 0.0223 (5) | 0.0399 (5) |

*Note*: The number shown in parentheses denotes the importance ranking of a variable for a given GSA index.

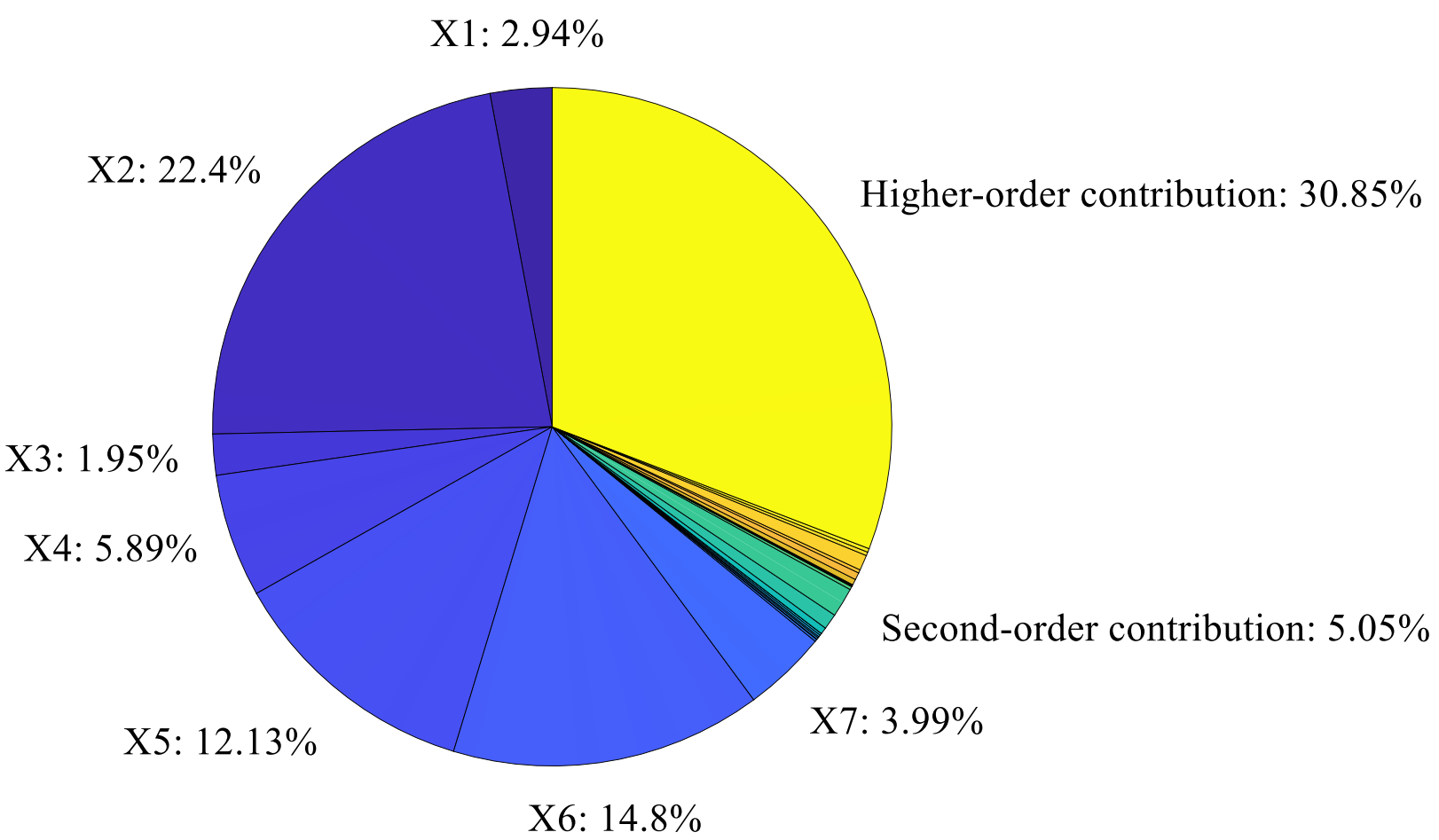

Fig. 4 Uncertainty decomposition results of the CRE-based measure for the probabilistic risk assessment model.

## 5 Application to a lifetime model of bearings

### 5.1 Bearing lifetime modification factor model

In this section, we show the superiority of the CRE-based measure in developing uncertainty reduction strategies when faced with highly-skewed distributions using a bearing lifetime modification factor model. Rolling bearings are critical components in rotating machinery, and their failure can lead to catastrophic consequences [39]. Therefore, accurately predicting bearing lifetime and reducing its uncertainty are of great importance. In practical applications, the bearing lifetime model based on the basic rating life $L_{10}$ has been widely used and proven to be effective, which can be expressed as:

$$L_{n,\text{modified}} = a_1 a_{ISO} L_{10} \tag{32}$$

where $L_{n,\text{modified}}$ denotes the lifetime with (1-$n$)% reliability after modification; $L_{10}$ represents the lifetime with 90 % reliability; $a_1$ is the modification factor for reliability related to $n$; and $a_{ISO}$ is the life modification factor which takes into account the fatigue stress limit of the bearing steel as well as the impacts of lubrication and contamination on bearing lifetime. To accurately determine $a_{ISO}$, a practical model was developed in ISO 281-2007, which can be expressed as [40]:

$$a_{ISO}=\begin{cases}0.1\left[1-\left(2.5671-\dfrac{2.2649}{k_0^{0.054381}}\right)^{0.83}\left(\dfrac{e_cC_u}{P}\right)^{1/3}\right]^{-9.3} & 0.1\le k_0<0.4\\ 0.1\left[1-\left(2.5671-\dfrac{1.9987}{k_0^{0.19087}}\right)^{0.83}\left(\dfrac{e_cC_u}{P}\right)^{1/3}\right]^{-9.3} & 0.4\le k_0<1\\ 0.1\left[1-\left(2.5671-\dfrac{1.9987}{k_0^{0.071739}}\right)^{0.83}\left(\dfrac{e_cC_u}{P}\right)^{1/3}\right]^{-9.3} & 1\le k_0<4\end{cases} \tag{33}$$

where $k_0$ denotes the viscosity ratio; $e_c$ is the contamination factor; $C_u$ is the fatigue load limit; and $P$ represents the dynamic equivalent load. The model is divided into three stages with the variation of $k_0$. Although the model is not strictly continuous in the mathematical sense, it can be considered as a continuous model in engineering applications.

In this case, we are concerned with how the uncertainty of these four input parameters influences the lifetime modification factor $a_{ISO}$ by Eq. (33). It is important to note that although the model used is relatively simple, it allows us to demonstrate the superiority of the CRE-based measure intuitively without interference from other potential effects.

In the absence of prior knowledge, the normal or uniform distribution is commonly used for uncertainty analysis [41, 42]. In this study, all input variables are assumed to follow normal distributions to simulate the actual operating conditions of bearings. Based on the conditions of bearings in starter-generators, as described in [43], the ranges of input variables are determined, and the corresponding distribution parameters are listed in Table 5. These parameters are configured to ensure that the majority of samples remain within the specified range. If a sampled value falls outside this range, it is resampled. Notably, the range and distribution of each input variable can be easily adjusted to accommodate different application scenarios if needed.

Table 5 Basic random variables and the distribution parameters for the bearing lifetime modification factor model.

| Random variable (unit) | Acceptable range | Distribution | Mean | Standard deviation |
|---|---|---|---|---|
| $k_0$ | 0.34 – 0.44 | Normal | 0.39 | 0.015 |
| $e_c$ | 0.5 - 1 | Normal | 0.75 | 0.08 |
| $C_u$ (kN) | 0.25 – 0.31 | Normal | 0.28 | 0.01 |
| $P$ (kN) | 9.5 – 13.5 | Normal | 11.5 | 0.6 |

### 5.2 Uncertainty importance analysis

Similarly, the CRE-based measure and the five GSA indices chosen in Section 4 are employed to analyze the uncertainty importance of each input variable. The hyper-parameter settings for the CRE-based measure remain unchanged. Table 6 presents the sensitivity results obtained using different GSA methods. Regarding the importance ranking, all moment-independent methods support that $e_c$ is the most influential variable, followed by $k_0$, whereas the Sobol index suggests that $k_0$ has the greatest impact.

Table 6 Global sensitivity results obtained by different GSA methods for the bearing lifetime modification factor model.

| Variable | $S$ | $S^T$ | $\delta$ | $\phi$ | $\eta$ | $\kappa$ |
|---|---|---|---|---|---|---|
| $k_0$ | 0.4686 (1) | 0.4695 (1) | 0.2342 (2) | 0.1007 (2) | 0.2732 (2) | 0.2639 (2) |
| $e_c$ | 0.3909 (2) | 0.3989 (2) | 0.2722 (1) | 0.1103 (1) | 0.3179 (1) | 0.2755 (1) |
| $C_u$ | 0.0415 (4) | 0.0444 (4) | 0.0704 (4) | 0.0096 (4) | 0.0236 (4) | 0.0289 (4) |
| $P$ | 0.0936 (3) | 0.0958 (3) | 0.1059 (3) | 0.0211 (3) | 0.0480 (3) | 0.0553 (3) |

*Note*: The number shown in parentheses denotes the importance ranking of a variable for a given GSA index.

To further interpret the results of different GSA methods, we separately fix $e_c$ and $k_0$ at their nominal values and compare the conditional PDF with the original one. The results are illustrated in Fig. 5.

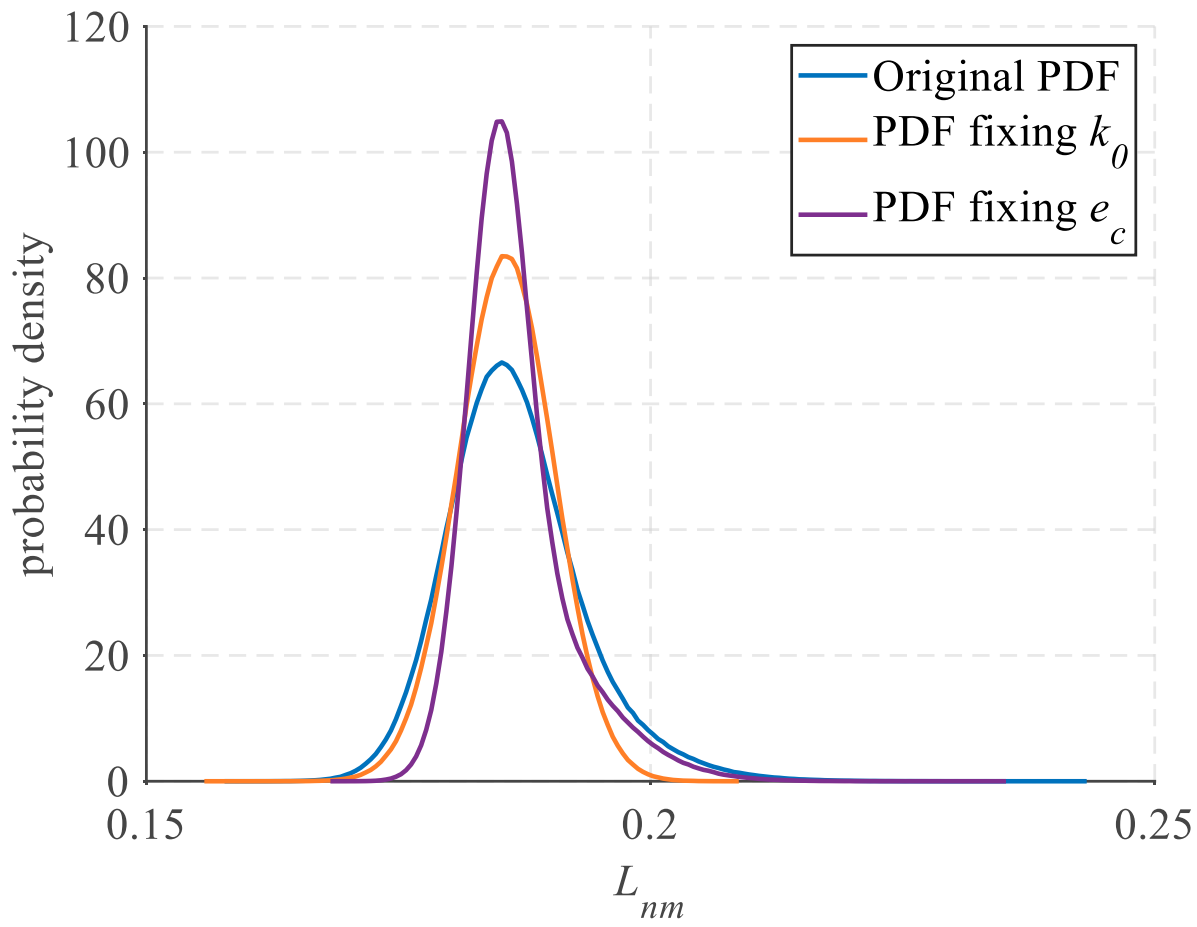

Fig. 5 PDF and conditional PDF of $a_{ISO}$.

From Fig. 5, it can be observed that the conditional PDF fixing $e_c$ is highly-skewed. In this context, fixing $e_c$ remains a large variance, which leads to its lower importance in the framework of Sobol index. In contrast, for the moment-independent measures, they are not influenced by the highly-skewed property of the PDF. As a result, these methods identify $e_c$ as the most important variable.

## 5.3 Uncertainty reduction strategy development considering costs

In real engineering scenarios, the cost of uncertainty reduction is significantly related to the uncertainty magnitude [27]. Therefore, it is necessary to quantify not only the uncertainty importance but also the uncertainty magnitude to develop an effective uncertainty reduction strategy. Among existing GSA measures, there are three methods capable of quantifying uncertainty magnitude: variance, differential entropy, and CRE, each corresponding to a specific uncertainty importance measure.

Table 7 shows the uncertainty magnitude of input and output variables obtained using different uncertainty measures. Notably, the results obtained by differential entropy include several negative values, making it unsuitable for representing actual uncertainty magnitude in practical applications. In contrast, both variance and CRE provide reasonable uncertainty magnitude for each variable.

Table 7 Uncertainty magnitude of input and output variables obtained by different uncertainty measures.

| Uncertainty measure | $k_0$ | $e_c$ | $C_u$ | $P$ | $a_{ISO}$ |
|---|---|---|---|---|---|
| Variance | 2.2025e-4 | 0.0064 | 1.0021e-4 | 0.3639 | 4.6812e-5 |
| Differential entropy | -2.7812 | -1.0941 | -3.1750 | 0.9236 | -3.5866 |
| CRE | 0.0134 | 0.0724 | 0.0091 | 0.5428 | 0.0065 |

Generally, when evaluating the cost of uncertainty reduction, analysts focus more on the relative uncertainty magnitude normalized by the mean (similar to the variation coefficient of a normal distribution). This is because the relative uncertainty magnitude directly influences the difficulty of reducing uncertainty. When the relative uncertainty magnitude is small, further reduction becomes increasingly challenging and costly.

In this paper, we define the cost function for uncertainty reduction as follows:

$$K_{i,j}\left(u_{i,j}\right) = K_0\left(\left(\frac{u_{reference,j}}{u_{i,j}}\right)^{\alpha} - 1\right), \quad u_{i,j} \le u_{reference,j} \tag{34}$$

where $u_{i,j}$ is the relative uncertainty magnitude of the $i^{th}$ input variable, and $j$ = 1, 2 denotes the method used: variance-based ($j$ = 1) or CRE-based ($j$ = 2); $u_{reference,j}$ is the reference relative uncertainty magnitude under cost $K_0$, typically derived from a specific distribution; $\alpha > 0$ denotes the nonlinearity in cost escalation with uncertainty reduction. For the normal distribution considered in this case, the relative uncertainty magnitude in the framework of CRE can be given as:

$$u_{i,2} = \frac{\mathcal{E}\left(X_i\right)}{\mu_i}, X_i \sim N\left(\mu_i, \sigma_i^2\right) \tag{35}$$

where $\mu_i$ and $\sigma_i$ are the mean value and standard deviation of the $i^{th}$ input variable, respectively. In the framework of variance, the relative uncertainty magnitude is expressed as:

$$u_{i,1} = \frac{\sigma_i}{\mu_i} \tag{36}$$

Then, given a specific $u_{reference,2}$ as 0.1, along with $K_0$ = 100 and $\alpha$ = 0.2, the uncertainty reduction costs for each input variable in the CRE-based framework can be calculated using Eq. (34), as listed in Table 8. From Table 8, the cost of reducing the uncertainty of $e_c$ is the lowest due to its relatively large uncertainty magnitude. On the other hand, $k_0$ and $C_u$ exhibit lower relative uncertainties, making further reduction more costly. Based on the results from Table 6 and Table 8, the CRE-based measure supports that controlling $e_c$ is the most effective and economical strategy.

Table 8 Relative uncertainty magnitude and uncertainty reduction cost of each input variable based on CRE.

| Uncertainty analysis items based on CRE | $k_0$ | $e_c$ | $C_u$ | $P$ |
|---|---|---|---|---|
| Uncertainty magnitude derived by Eq. (5) | 0.0134 | 0.0724 | 0.0091 | 0.5428 |
| Relative uncertainty magnitude derived by Eq. (35) | 0.0348 | 0.0964 | 0.0322 | 0.0471 |
| Uncertainty reduction cost derived by Eq. (34) | 23.5 | 0.736 | 25.4 | 16.3 |

Next, the uncertainty reduction cost of each input variable should by computed in the framework of variance for comparison. Notably, according to Eq. (34), it can be demonstrated that for certain distributions, the uncertainty reduction cost remains identical whether measured using variance or CRE. Specifically, for commonly encountered uniform and normal distributions, Eqs. (9) and (12) indicate a linear relationship between CRE and standard deviation. Hence, given $u_{reference,2}$ as 0.1 for a specific normal distribution, $u_{reference,1}$ can be derived as 0.1107 according to Eq. (12). As a result, the uncertainty reduction costs computed via Eq. (34) are equivalent for both methods.

In this bearing case, since all input variables follow a normal distribution, the uncertainty reduction costs derived by CRE and variance are equal. Therefore, the uncertainty reduction strategy is completely governed by the uncertainty importance rankings of input variables in Table 6. Then, an intuitive yet essential question arises: which GSA measure should be chosen as the foundation for decision making? As discussed in [44, 45], the choice of a GSA method should align with the analyst's objectives or the specific requirements of the study. For instance, moment-independent approaches should be preferable and the uncertainty of variable $e_c$ should be prioritized for control if analysts are concerned with changes in the output distribution. Conversely, the variance-based method may be more appropriate and the uncertainty of variable $k_0$ should be controlled as a priority if the measure of central tendency is emphasized.

## 6 Conclusion

In addition to quantifying uncertainty importance, uncertainty magnitude quantification is also essential in developing effective uncertainty reduction strategies. However, there is a lack of moment-independent GSA indices to achieve this purpose, resulting in challenges on providing effective guidance for developing uncertainty reduction strategies when faced with highly-skewed distributions. Motivated by this problem, a new moment-independent uncertainty importance measure based on CRE is proposed. Uncertainty magnitude and uncertainty importance quantification can be achieved simultaneously by the proposed measure. Corresponding numerical implementations of the proposed index are devised and verified. Then, two numerical examples and an engineering case are conducted to demonstrate the effectiveness of the proposed index. Some conclusions could be drawn as follows:

- Numerical examples show that the CRE-based importance measure can give effective importance rankings of uncertainty importance as other moment-independent methods and is capable of capturing the actual interaction contributions.
- The engineering case indicates that the proposed measure gives a different uncertainty reduction recommendation compared to the Sobol index, which reveals its superiority in handling highly-skewed distributions. Compared with other moment-independent GSA methods, the proposed measure is able to present a reasonable uncertainty magnitude quantification, which can help analysts balance the uncertainty reduction cost in developing uncertainty reduction strategies.

Consequently, it is demonstrated that the proposed CRE-based uncertainty importance measure can serve as a reliable tool for providing guidance on uncertainty reduction. Future research will focus on applying the CRE-based measure to analyze correlated variables.

## Acknowledgements

This work was supported by the National Natural Science Foundation of China [grant number 51775020], the Science Challenge Project [grant number. TZ2018007], the National Natural Science Foundation of China [grant numbers 62073009].

## Appendix A

### A.1. Proof of Property 1

***Property 1***. $0 \le \kappa_i \le 1$.

*Proof*: It can be easily seen that $\kappa_i$ is nonnegative because of *Theorem 2*. Then, we need to check that $\kappa_i \le 1$, i.e. $E_{X_i}\left[\mathcal{E}\left(Y|X_i = x_i\right)\right] \ge 0$. Recall the definition of a sigma field from probability theory: a sigma field $\mathcal{F}$ is a class of subsets containing the empty set and closed under compliments and countable unions. Then, $\mathcal{E}\left(Y|X_i = x_i\right)$ can be measurable with respect to a specific sigma field $\mathcal{F}_i$. From *Theorem 1*, we have $\mathcal{E}\left(\mathcal{F}_i\right) \ge 0$. □

### A.2. Proof of Property 3

***Property 3***. $0 \le \kappa_{ij} \le 1$.

*Proof*: First, we prove that $\kappa_{ij} \ge 0$. For this purpose, we introduce the theory of the partial information decomposition (PID) [46]. According to the PID, the information interaction $I_{\mathcal{E}}\left(X_i, X_j; Y\right)$ can be decomposed into four non-negative parts: the isolated contributions of the two variables, their redundant contribution, and their synergistic contribution. Since $X_i$ and $X_j$ are independent, their redundant contribution is zero. Moreover, $I_{\mathcal{E}}\left(X_i; Y\right)$ and $I_{\mathcal{E}}\left(X_j; Y\right)$ represent their isolated contribution, respectively. Since the synergistic contribution is non-negative, we derive that $I_{\mathcal{E}}\left(X_i, X_j; Y\right) - I_{\mathcal{E}}\left(X_i; Y\right) + I_{\mathcal{E}}\left(X_j; Y\right) \ge 0$. Therefore, we have $\kappa_{ij} \ge 0$.

Then, we prove that $\kappa_{ij} \le 1$. According to Eqs. (16) and (17), Eq. (18) can be rewritten as:

$$\kappa_{ij} = \frac{E_{X_i}\left[\mathcal{E}\left(Y|X_i = x_i\right)\right] + E_{X_j}\left[\mathcal{E}\left(Y|X_j = x_j\right)\right] - E_{X_i,X_j}\left[\mathcal{E}\left(Y|X_i = x_i, X_j = x_j\right)\right] - \mathcal{E}\left(Y\right)}{\mathcal{E}\left(Y\right)}. \quad (37)$$

Then, we need to verify that

$$2\mathcal{E}\left(Y\right) + E_{X_i,X_j}\left[\mathcal{E}\left(Y|X_i = x_i, X_j = x_j\right)\right] \ge E_{X_i}\left[\mathcal{E}\left(Y|X_i = x_i\right)\right] + E_{X_j}\left[\mathcal{E}\left(Y|X_j = x_j\right)\right]. \quad (38)$$

If $X_i$ and $X_j$ are both independent of $Y$, then it can be derived that

$$\begin{aligned} &2\mathcal{E}\left(Y\right) + E_{X_i,X_j}\left[\mathcal{E}\left(Y|X_i = x_i, X_j = x_j\right)\right] = 3\mathcal{E}\left(Y\right) \\ &\ge 2\mathcal{E}\left(Y\right) = E_{X_i}\left[\mathcal{E}\left(Y|X_i = x_i\right)\right] + E_{X_j}\left[\mathcal{E}\left(Y|X_j = x_j\right)\right]. \end{aligned} \quad (39)$$

On the other hand, if $X_i$ is not independent of $Y$ (same for $X_j$), we have:

$$2\mathcal{E}(Y)+E_{X_i,X_j}\left[\mathcal{E}\left(Y\middle|X_i=x_i,X_j=x_j\right)\right]=2\mathcal{E}(Y)+E_{X_j}\left[\mathcal{E}\left(Y\middle|X_j=x_j\right)\right]$$
$$\geq\mathcal{E}(Y)+E_{X_j}\left[\mathcal{E}\left(Y\middle|X_j=x_j\right)\right]=E_{X_i}\left[\mathcal{E}\left(Y\middle|X_i=x_i\right)\right]+E_{X_j}\left[\mathcal{E}\left(Y\middle|X_j=x_j\right)\right] \tag{40}$$

□

A.3. Proof of Property 4

***Property 4***. $\sum_{i=1}^{n}\kappa_i+\sum_{1\leq i<j\leq n}\kappa_{ij}+\ldots+\kappa_{12\ldots n}=1$.

*Proof*: By using Eq. (19), proving the above equation amounts to proving that $I_{\mathcal{E}}\left(X_1,X_2,...,X_n;Y\right)=\mathcal{E}(Y)$. Furthermore, based on Eq. (17), we simply need to verify that $E_{X_i,X_j,...,X_n}\left[\mathcal{E}\left(Y\middle|X_i=x_i,X_j=x_j,...,X_n=x_n\right)\right]=0$. According to Eq. (13), $Y$ is measurable to a sigma field formed by $X_1,X_2,...,X_n$. Subsequently, by using *Theorem 3*, we can derive that $E_{X_i,X_j,...,X_n}\left[\mathcal{E}\left(Y\middle|X_i=x_i,X_j=x_j,...,X_n=x_n\right)\right]=0$. □

## Appendix B

B.1 Validation of the algorithm for estimating CRE

In this subsection, we check the effectiveness of the proposed algorithm for estimating CRE in Section 3.4.1. We investigate a random variable $X$ that follows an exponential distribution, i.e. $X\sim$ Exp (0.5). According to Eq. (7), we can derive that $\mathcal{E}(X)=2$. Then, we generate a series of samples of $X$ and estimate $\mathcal{E}(X)$ based on the proposed algorithm. The sample size ranges from 100 to 20000. The convergence and computation time of the proposed algorithm with different sample size are investigated as shown in Fig. 6.

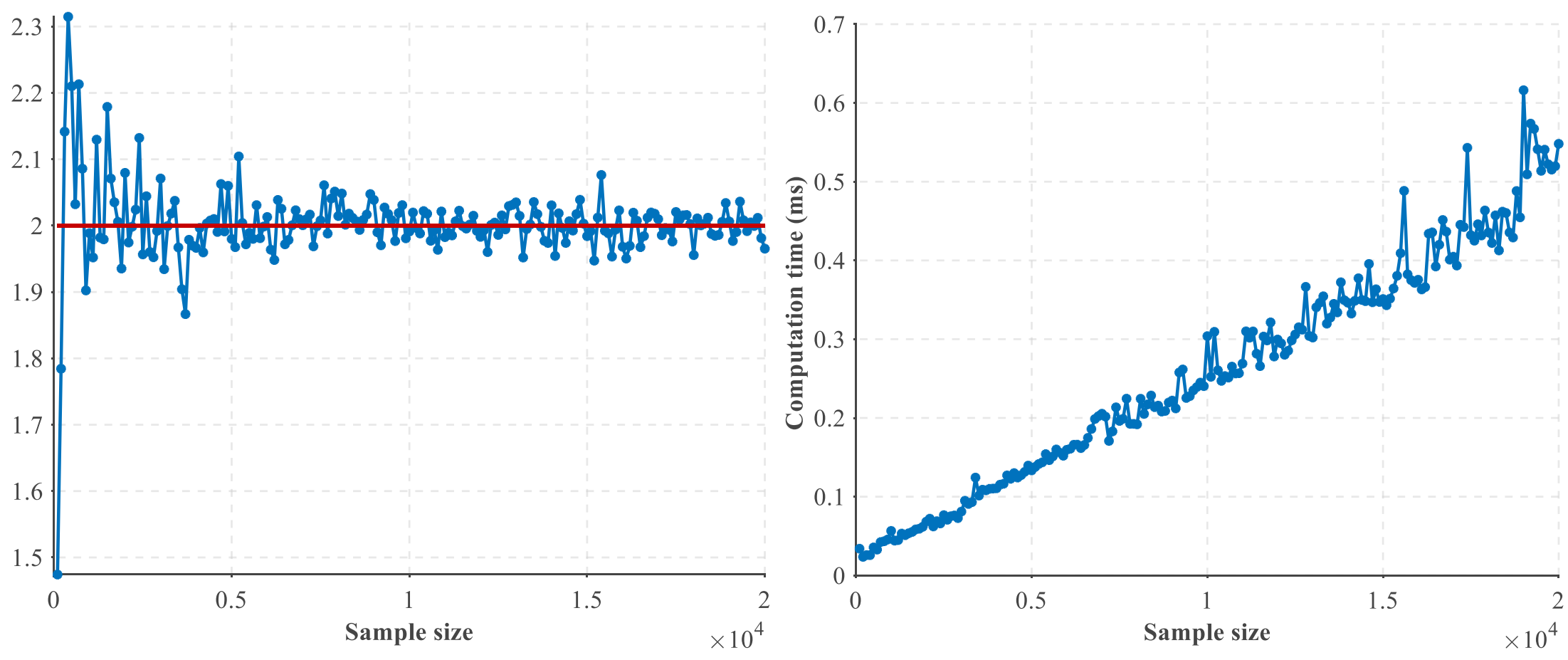

(a) (b)

Fig. 6 The convergence and computation time of the proposed algorithm for estimating CRE with different sample size.

It can be seen from Fig. 6 (a) that the proposed algorithm has a high accuracy with a small sample size. When the sample size is 1000, the corresponding estimation is 1.989. From Fig. 6 (b), the computational cost required by the algorithm is small and the computational complexity increases linearly with the sample size, which indicates the superiority of the CDF-based GSA index in computation time. When the sample size is 20000, the computation time is only 0.54 ms. The above results show that the proposed algorithm can provide fast and accurate estimation of CRE.

B.2 Validation of the algorithm for estimating conditional CRE considering single variable

In this subsection, we check the effectiveness of the proposed algorithm for estimating conditional CRE considering single variable in Section 3.4.2. Consider the following function:

$$Y = X_1 + X_2 \tag{41}$$

where $X_1 \sim$ Exp (0.5) and $X_2 \sim N$ (40, 4). In this case, when we fix $X_2$, the uncertainty of $Y$ will be entirely determined by $X_1$, i.e. $\mathcal{E}(Y|X_2) = \mathcal{E}(X_1) = 2$. Then, we generate a series of samples of $X_1$ and $X_2$, and estimate $\mathcal{E}(Y|X_2)$ based on the proposed algorithm. The hyper-parameter $m$ is set as 500. The sample size ranges from 500 to 20000. The convergence and computation time of the proposed algorithm with different sample size are investigated as shown in Fig. 7.

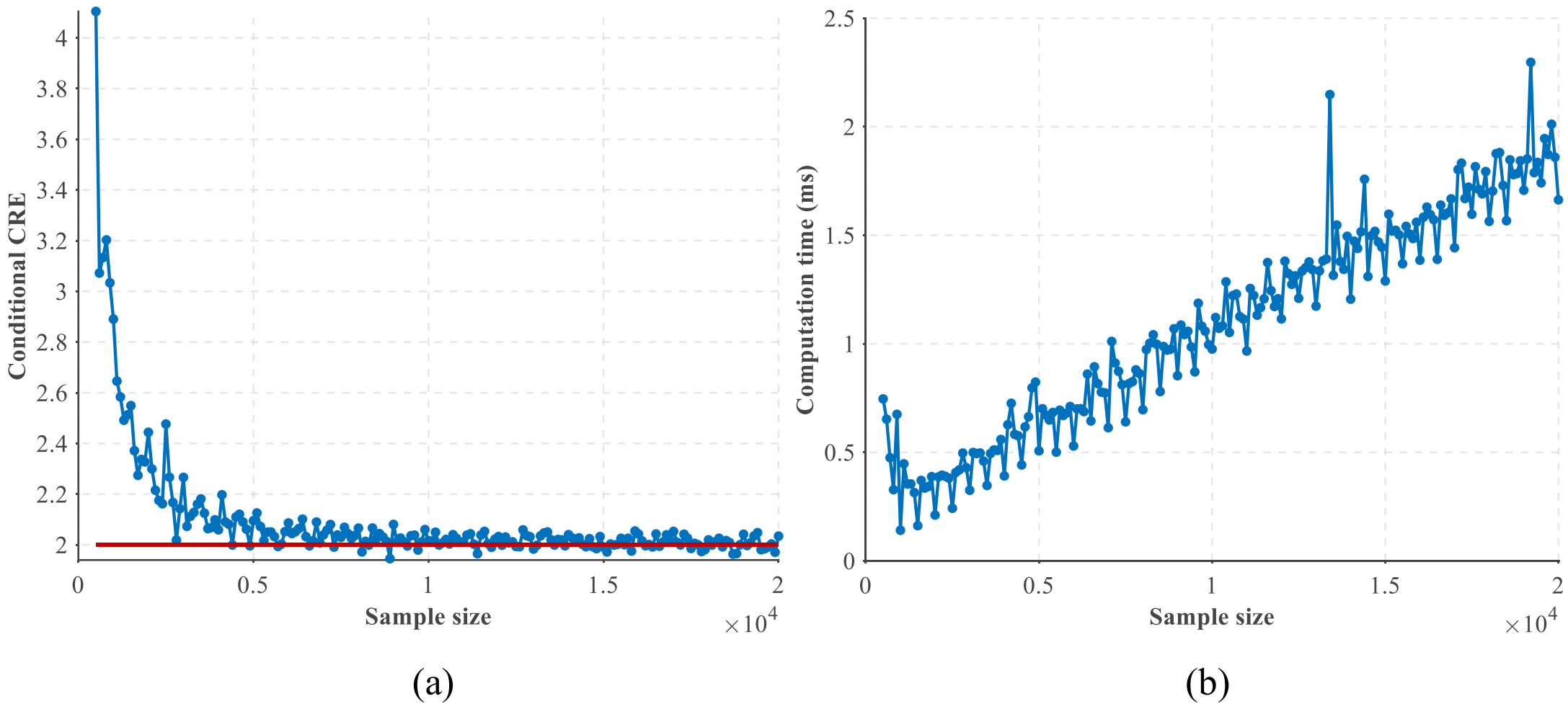

Fig. 7 The convergence and computation time of the proposed algorithm for estimating conditional CRE considering single variable with different sample size.

From Fig. 7 (a), the proposed algorithm gets convergence over 10000 samples. When the sample size is 5000, the corresponding estimation is 2.095. From Fig. 7 (b), the computational cost required by the algorithm is small and the computational complexity increases linearly with the sample size, which indicates the superiority of the CDF-based GSA index in computation time. When the sample size is 20000, the computation time is only 1.66 ms. The above results show that the proposed algorithm can provide fast and accurate estimation of conditional CRE considering single variable.

B.3 Validation of the algorithm for estimating conditional CRE considering two variables

In this subsection, we check the effectiveness of the proposed algorithm for estimating conditional CRE considering two variables in Section 3.4.3. Consider the following function:

$$Y = X_1 + X_2 + X_3 \tag{42}$$

where $X_1 \sim$ Exp (0.5), $X_2 \sim$ Exp (0.1) and $X_3 \sim N$ (40,4). In this case, when we fix $X_2$ and $X_3$, the uncertainty of $Y$ will be entirely determined by $X_1$, i.e. $\mathcal{E}(Y|X_2, X_3) = \mathcal{E}(X_1) = 2$. Then, we generate a series of samples of $X_1$, $X_2$ and $X_3$, and estimate $\mathcal{E}(Y|X_2, X_3)$ based on the proposed algorithm. The hyper-parameter $I$ and $J$ are set as 20. The sample size ranges from 100 to 20000. The convergence and computation time of the proposed algorithm with different sample size are investigated as shown in Fig. 8.

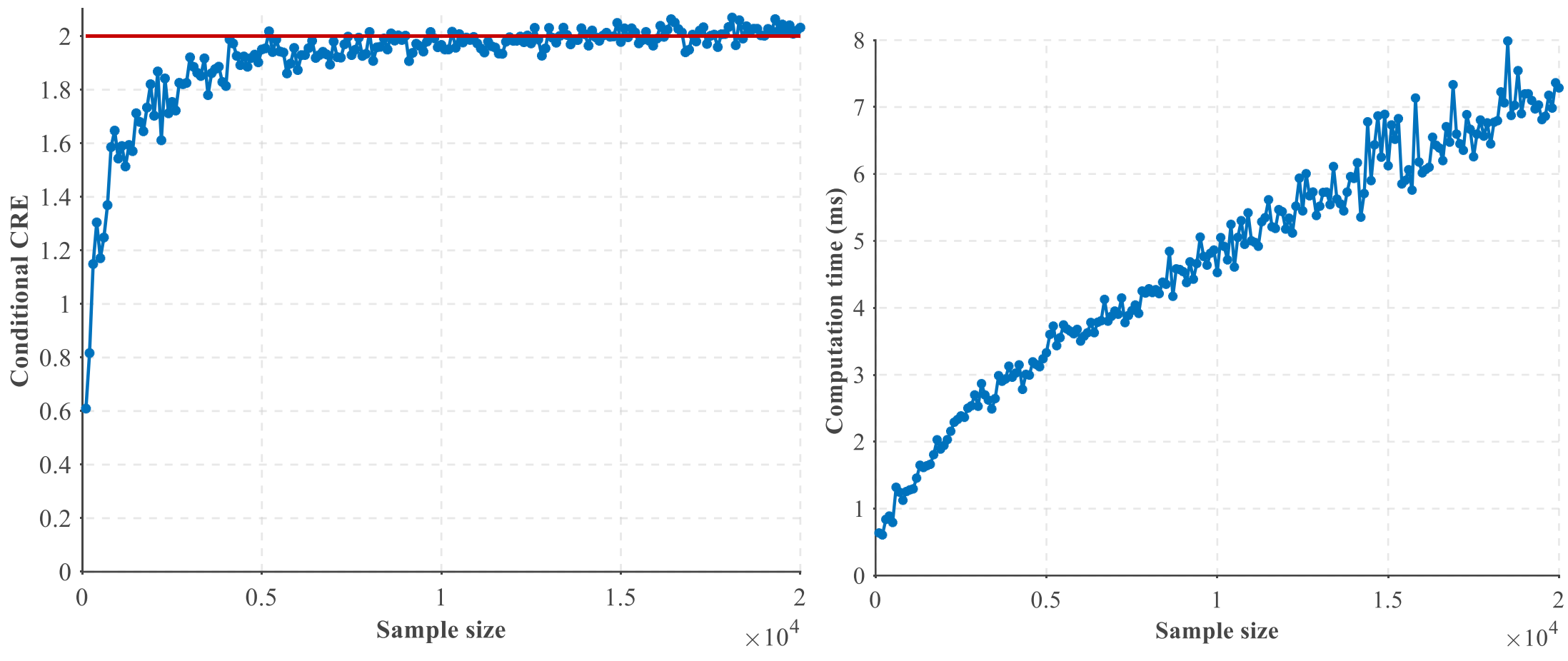

(a) (b)

Fig. 8 The convergence and computation time of the proposed algorithm for estimating conditional CRE considering two variables with different sample size.

As illustrated in Fig. 8 (a), the proposed algorithm gets convergence over 10000 samples. When the sample size is 5000, the corresponding estimation is 1.95. From Fig. 8 (b), the computational cost required by the algorithm is small and the computational complexity increases linearly with the sample size, which indicates the superiority of the CDF-based GSA index in computation time. When the sample size is 20000, the computation time is only 7.32 ms. The above results show that the proposed algorithm can provide fast and accurate estimation of conditional CRE considering two variables.

## Appendix C

### C.1 Sensitivity analysis of hyber-parameter *m*

In this subsection, we explore the influence of hyber-parameter *m* on the estimation of conditional CRE considering single variable. We consider the Eq. (41) in Appendix B.2 and stay the parameter settings unchanged. The sample size ranges from 1000 to 50000. The convergence of the proposed algorithm with different sample size under different hyber-parameter *m* is investigated as shown in Fig. 9.

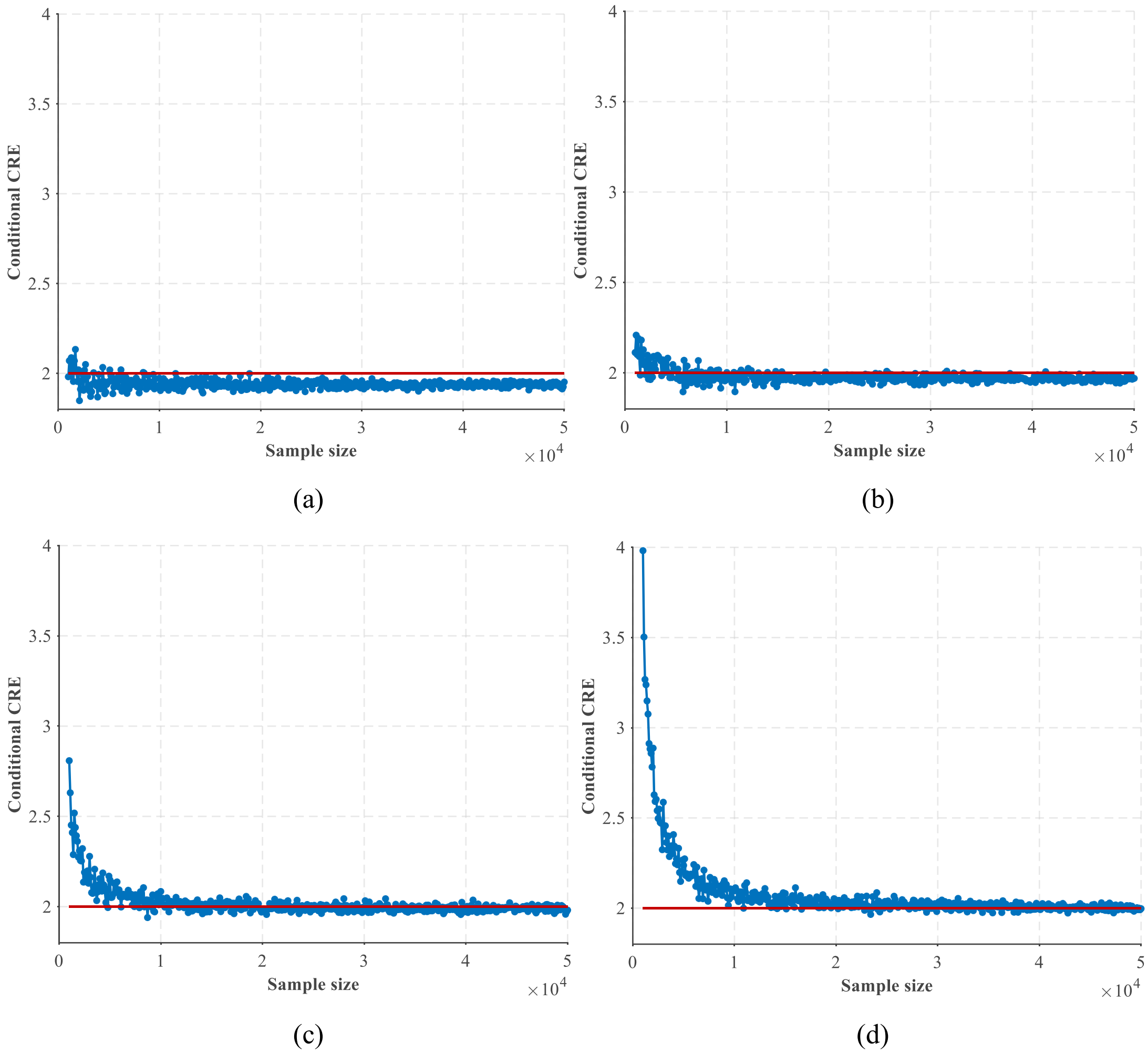

Fig. 9 The convergence of the proposed algorithm for estimating conditional CRE considering single variable with different sample size: (a) $m = 100$ (b) $m = 200$ (c) $m = 500$ (d) $m = 1000$.

From Fig. 9, there is an estimation bias if $m$ is too small. When $m = 100$, the estimation converges to nearly 1.94. When $m = 200$, the estimation converges to around 1.97. Then, when $m = 500$, the estimation converges to 2 over 10000 samples. However, if $m$ is too large, like $m = 1000$, the algorithm requires more sample size to reach convergence. As a result, $m = 500$ is chosen for the case studies in this paper.

C.2 Sensitivity analysis of hyber-parameters $I$ and $J$

In this subsection, we explore the influence of hyber-parameter $I$ and $J$ on the estimation of conditional CRE considering two variables. We consider the Eq. (42) in Appendix B.3 and stay the parameter settings unchanged. The sample size ranges from 100 to 20000. The convergence of the proposed algorithm with different sample size under different hyber-parameter $I$ and $J$ is investigated as shown in Fig. 10.

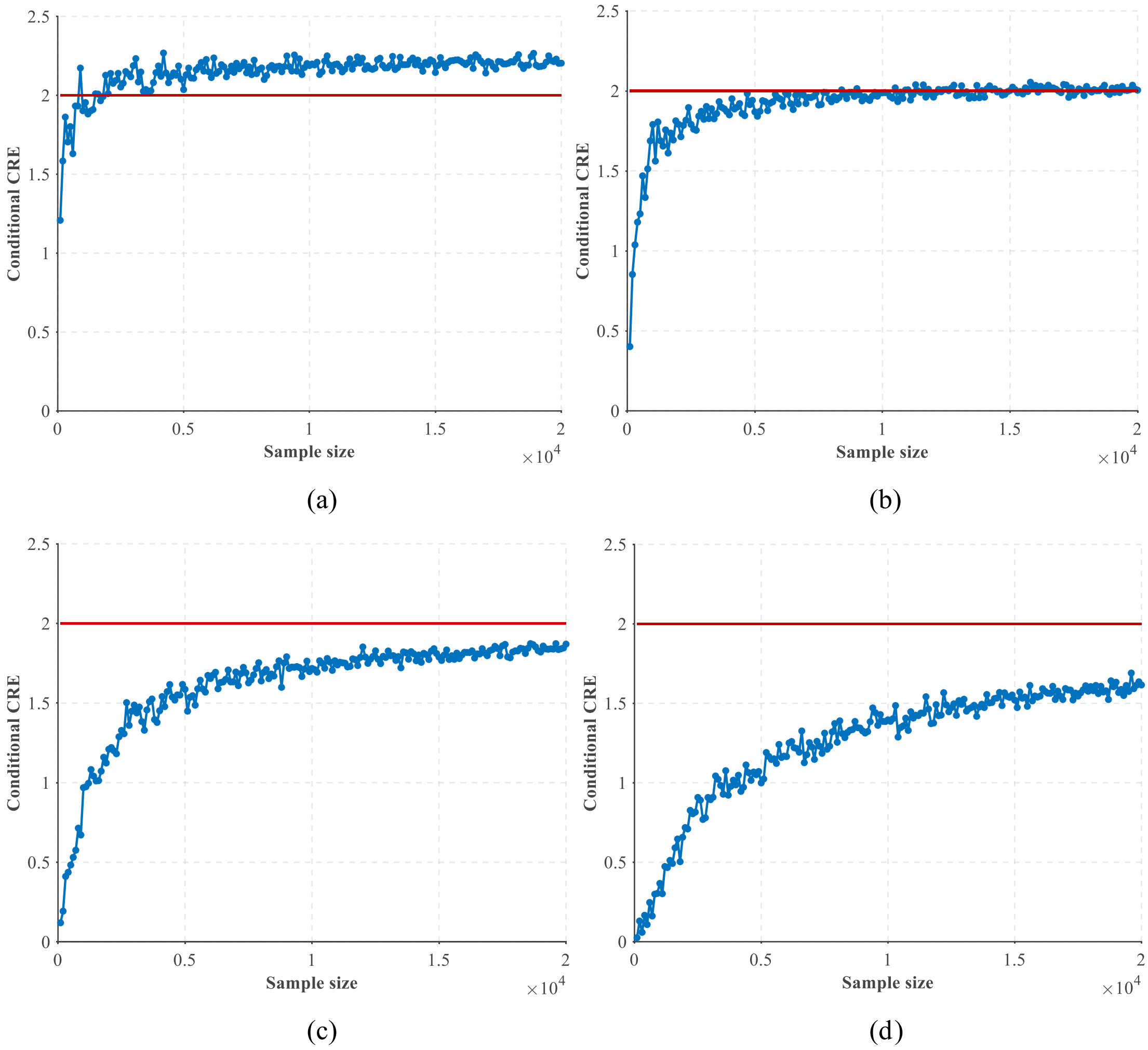

Fig. 10 The convergence of the proposed algorithm for estimating conditional CRE considering two variables with different sample size: (a) $I = J = 10$ (b) $I = J = 20$ (c) $I = J = 50$ (d) $I = J = 100$.

As illustrated in Fig. 10, there is an estimation bias if $m$ is too small. When $I = J = 10$, the estimation converges to nearly 2.2. Then, when $I = J = 20$, the estimation converges to 2 over 10000 samples. However, if $m$ is too large, the algorithm requires more sample size to reach convergence. For $I = J = 50$ and $I = J = 100$, they fail to reach convergence under 20000 samples. As a result, $I = J = 20$ is chosen for the case studies in this paper.